\documentclass[11pt]{article}
\usepackage{epsfig}

\normalbaselines
\newcounter{eqnletter}[equation]
\catcode`\@=11
\@addtoreset{equation}{section}   
\catcode`\@=12

\newcommand{\be}{\begin{eqnarray}}
\newcommand{\ee}{\end{eqnarray}}

\def\ll#1{\left#1}

\def\fr{\frac{1}{2}}

\def\mref#1{(\ref{#1})}

\def\bd{\begin{displaymath}}
\def\ed{\end{displaymath}}

\def\ba#1{\begin{array}{#1}}
\def\ea{\end{array}}
\def\nn{\nonumber}

\begin{document}

\pagestyle{empty}

\begin{center}

{\LARGE\bf CHANGE OF VARIABLES AS BOREL RESUMMATION OF
SEMICLASSICAL SERIES\\[0.5cm]}

\vskip 60pt

{\large {\bf Stefan Giller{$\dag$}\footnote{Supported by KBN
2PO3B 07610} and Piotr Milczarski{$\ddag$}\footnote{Supported by
the {\L}\'od\'z University Grant No 580} }}

\vskip 18pt

Theoretical Physics Department II, University of {\L}\'od\'z,\\
Pomorska 149/153, 90-236 {\L}\'od\'z, Poland\\ e-mail: $\dag$
sgiller@krysia.uni.lodz.pl \\ $\ddag$
jezykmil@krysia.uni.lodz.pl
\end{center}
\vspace{3 cm}
\begin{abstract}

It is shown that a change of variable in 1-dim
Schr\"odinger equation applied to the Borel summable
fundamental solutions \cite{8,9} is equivalent to Borel
resummation of the fundamental solutions multiplied by suitably
chosen $\hbar$-dependent constant. This explains why change of
variable can improve JWKB formulae \cite{11}.
It is shown also that a change of variable
alone cannot provide us with the exact JWKB formulae.
\vskip 36pt

PACS number(s): 03.65.-W , 03.65.Sq , 02.30.Lt , 02.30.Mv


\end{abstract}

\newpage

\pagestyle{plain}

\setcounter{page}{1}

\section{Introduction }
\vskip 12pt

A change of variable in the 1-dim Schr\"odinger equation
(1-DSE) is one of the basic technic used to solve 1-dim problems
(see \cite{12} for example). In the context of semiclassical (JWKB)
approximation the procedure is in fact a main ingredient of
Fr\"oman and Fr\"oman (F-F) approach to 1-DSE \cite{3,4}
with an aim of getting improved JWKB quantization formulae \cite{4,5,6}.
Sometimes, a suitable change of variable provides us with
JWKB-like formulae solving the problem of energy spectra even
exactly \cite{4}. No doubts, however, that the latter
possibility depends totally on a potential considered and
changing variable plays in such cases only an auxiliary role
\cite{11}.  

  A change of variables is
also an essential ingredient of a more general approach to the
semiclassical approximations formulated by Maslov and his
collaborators [13]. In the context of the latter approach the
change-of-variable procedure is an inherent part of the
continuation procedure of semiclassical series defined
originally in some domain of the configuration space to another
domain of the space. The relevant variable transformations used
in the Maslov method are the canonical ones (in a sens of
classical mechanics). A relation of the Maslov method to the one
applied in this paper is discussed in Appendix. It is argued
there however that using fundamental solutions as we do in our
paper is equivalent to the method of Maslov \underline{et al} in the
semiclassical regime of the considered 1-dim problems but has
many advantages over the Maslov procedure in the remaining of
our investigations. In particular the problem of Borel
resummation central for our paper cannot be put and considered
properly ignoring the existence of the fundamental solutions and
their properties. After all the method of Maslov \underline{et al} is purely
asymptotic from the very begining and any problem of resummation
of the semiclassical series used in the method has not been
considered as yet.  

An improvement of the standard JWKB formulae achieved by the
changing variable procedure appears as corrections having
typically forms of additional $\hbar$-dependent term in emerging
effective potentials \cite{1,4,5,6}.  Since in all these cases of 
variable changing the standard JWKB formulae can be easily restored
simply by $\hbar$-expansions of the improved ones the latter
seems to be a kind of some hidden resummation of a part (in the
case of improvements only) or a full (when exact formulae
emerge) standard semiclassical expansion corresponding to
considered cases.

It is the aim of this paper to show that indeed 
this hidden resummation mentioned above
really takes place and that a class of applied changes of
variable in 1-DSE results as the Borel resummation of suitably
chosen standard
semiclassical solutions to SE multiplied by appropriately chosen
$\hbar$-dependent constants which can always be
attached to any of such semiclassical solution.  

As it has been shown by Milczarski and Giller \cite{7} (see also \cite{8})
such specific Borel summable solutions to SE are provided
for meromorphic potentials by the F-F construction
\cite{3} in the form of so called fundamental solutions (FS)\cite{8,9}. These
 are the only solutions with the Borel summability  property among all the 
F-F-like solutions \cite{7}. Despite their rareness the FS's when collected
into a full set of them allow us to solve any 1-dim problem \cite{8,9}
(see also a discussion below).

The paper is organized as follows.  

In the next section the fundamental solutions and their use are
recalled.  

In Sec.3 the standard semiclassical expansions and
their properties are reconsidered.  

In Sec.4 the Borel
resummation aspects of a change-of-variable operation are
discussed.  

In Sec.5 the impossibility of achieving the exact
JWKB formulae by a change-of-variable operation only is
discussed.  

We conclude with Sec.6. 

\section{ Fundamental solutions }

A standard way to introduce FS's is a construction of
a Stokes graph (SG) \cite{7,8,9} for a given (meromorphic)
potential $V(x)$.

SG consists of Stokes lines (SL) emerging from roots (turning
points) of the equation:
\begin{eqnarray}
V(x) + \hbar^2 \delta(x) = E
\label{1}
\end{eqnarray}
with $E$ as energy as well as from simple poles of the
considered potential $V(x)$. 

The presence and role of the $\delta$-term in (\ref{1}) is explained below.
It contributes to (\ref{1}) only when $V(x)$ contains simple and second order
poles. The $\delta$-term is constructed totally from these poles.

The points of SL's satisfy one of the following equations: 
\begin{eqnarray}
\Re \int_{x_{i}}^{x} \sqrt{V(y) + {\hbar^2}\delta(y)- E}dy = 0
\label{2}
\end{eqnarray}
with $x_{i}$ being a root of (\ref{1}) or a simple pole of $V(x)$.

SL's which are not closed end at these points of the $x$-plane 
(i.e. have the latter as the boundaries) for which an action
integral in (\ref{2}) becomes infinite. Of course such points
are singular for the potential $V(x)$ and can be finite poles,
higher than the simple ones, or poles of $V(x)$ lying at the infinity.

Each such a singularity $x_{0}$ of $V(x)$ defines a domain called a sector.
This is the connected domain of the x-plane bounded by the SL's and $x_0$
itself with the latter point being also a boundary for the SL's or being 
an isolated boundary point of the sector (as it is in the case of the second
order pole).

In each sector the LHS in (\ref{2}) is only positive or negative.

Consider now the Schr\"odinger equation:
\begin{eqnarray}
\Psi^{\prime\prime}(x) - \hbar^{-2} q(x) \Psi(x) = 0
\label{7}
\end{eqnarray}
where $q(x)=V(x)-E$ (we have put the mass m in (\ref{7}) to be equal 
to $1/2$). 

Following Fr\"oman and Fr\"oman one can define in each sector $k$ having 
$x_0$ at its boundary a solution of the form: 
\begin{eqnarray}
\Psi_{k}(x) = \tilde{q}^{-\frac{1}{4}}(x){\cdot}
e^{\frac{\sigma}{\hbar} W(x)}{\cdot}{\chi_{k}(x)} &
& k = 1,2,\ldots
\label{3}
\end{eqnarray}
where:
\begin{eqnarray}
\chi_{k}(x) = 1 + \sum_{n{\geq}1}
\left( -\frac{\sigma \hbar}{2} \right)^{n} \int_{x_{0}}^{x}d{\xi_{1}}
\int_{x_{0}}^{\xi_{1}}d{\xi_{2}} \ldots 
\int_{x_{0}}^{\xi_{n-1}}d{\xi_{n}}
\omega(\xi_{1})\omega(\xi_{2}) \ldots \omega(\xi_{n}) 
\label{4}
\end{eqnarray}
\begin{eqnarray*}
{\times} \left( 1 -
e^{-\frac{2\sigma}{\hbar}{(W(x)-W(\xi_{1}))}} \right)
\left(1 - e^{-\frac{2\sigma}{\hbar}{(W(\xi_{1})-W(\xi_{2}))}} \right)
\cdots
\left(1 - e^{-\frac{2\sigma}{\hbar}{(W(\xi_{n-1})-W(\xi_{n}))}} \right)
\end{eqnarray*}
with
\begin{eqnarray}
\omega(x) = \frac{\delta(x)}{\tilde{q}^{\frac{1}{2}}(x)} - 
{\frac{1}{4}}{\frac{\tilde{q}^{\prime\prime}}{\tilde{q}^{\frac{3}{2}}(x)}} +
{\frac{5}{16}}{\frac{\tilde{q}^{\prime 2}}{\tilde{q}^{\frac{5}{2}}(x)}}
\label{5}
\end{eqnarray}
and
\begin{eqnarray}
W(x,E) = \int_{x_{i}}^{x} \sqrt{\tilde{q}(\xi,E)}d\xi
\label{6}
\end{eqnarray}
\begin{eqnarray*}
\tilde{q}(x,E) = V(x) +\hbar^2 \delta(x) - E
\end{eqnarray*}

In (\ref{3}) and (\ref{4}) a sign $\sigma (= \pm 1)$ and an
integration path are chosen in such a way to have: 
\begin{eqnarray}
\sigma \Re \left(W(\xi_{j}) - W(\xi_{j+1}) \right) \leq 0
\label{8}
\end{eqnarray}
for any ordered pair of integration variables (with $\xi_{0} =
x$). Such a path of integration is then called canonical.  

The term $\delta(x)$ appearing in (\ref{5}) and in (\ref{6}) 
is necessary to ensure all the integrals in (\ref{4}) to
converge when $x_0$ is a first or a second order pole of $V(x)$
or when solutions (\ref{3}) are to be continued to such poles.
Each such a pole $x_0$ demands a contribution to $\delta(x)$ 
of the form $(2(x-x_0))^{-2}$ so that $\delta(x)$ collects all 
of them and its final form depends of course on the corresponding
singular structure of $V(x)$.

Note that the effect of introducing the $\delta$-term is completely
equivalent to making some change of variable in the SE, 
a possibility which in this context shall, however, not be 
discussed in the paper.  

In a domain $D_{k}$ of the $x$-plane where the condition
(\ref{8}) is satisfied (so called canonical domain) the series
in (\ref{4}) defining $\chi_{k}$ is uniformly convergent.
$\chi_{k}$ itself satisfies the following initial conditions:
\begin{eqnarray}
\chi_{k}(x_{0}) = 1 & \mbox{and} & \chi_{k}^{\prime}(x_{0}) = 0
\label{9}
\end{eqnarray}
corresponding to the equation:
\begin{eqnarray}
\chi_{k}(x) = 1 - 
\frac{\sigma{\hbar}}{2}\int_{x_{0}}^{x}dy{\omega(y)}\chi_{k} - 
\frac{\sigma{\hbar}}{2}\tilde{q}^{-\frac{1}{2}}(x)\chi_{k}^{\prime}(x) 
\label{10}
\end{eqnarray}
this function has to obey as a consequence of SE (\ref{7}) and
the initial conditions (\ref{9}).  

In the canonical domain $D_{k}$ and the sector $S_{k} (\subset
D_{k})$ where the solution (\ref{3}) is defined the latter has
two following basic properties: 

$1^{0}$ It can be expanded in $D_{k}$ into a standard
semiclassical series obtained by iterating Eq.(\ref{10}) and
taking into account the initial conditions (\ref{9});

$2^{0}$ The emerging semiclassical series is Borel summable in
$S_{k}$ to the solution itself.  

The solutions (\ref{3}) defined in the above way are known as
the fundamental ones \cite{8,9}. They are pairwise independent
and collected into a full set of them they allow to solve
\underline{any} one-dimensional problem. They are distinguished by the
property $2^{0}$ above i.e. they are the unique solutions to SE
with this property \cite{7}.

\section{Standard semiclassical expansions}

By a standard semiclassical expansion for $\chi$ we
mean the following series:
\begin{eqnarray*}
\chi(x) \sim C(\hbar)\sum_{n\geq{0}} 
\left(-\frac{\sigma{\hbar}}{2} \right)^{n}
\chi_{n}(x)
\end{eqnarray*}
\begin{eqnarray*}
\chi_{0}(x) = 1
\end{eqnarray*}
\begin{eqnarray}
\chi_{n}(x) = \int_{x_{0}}^{x}d\xi_{n}\tilde{D}(\xi_{n}) 
\times
\label{11}
\end{eqnarray}
\begin{eqnarray*}
\times \int_{x_{0}}^{\xi_{n}}d\xi_{n-1}\tilde{D}(\xi_{n-1}) 
\ldots \int_{x_0}^{\xi_3} d\xi_2 \tilde{D}(\xi_2) 
\int_{x_{0}}^{x}d\xi_{1}\left( \tilde{q}^{-\frac{1}{4}}(\xi_{1})
\left( \tilde{q}^{-\frac{1}{4}}(\xi_{1}) \right)^{\prime{\prime}} +
\tilde{q}^{-\frac{1}{2}}(\xi_{1})\delta(\xi_1)\right) 
\end{eqnarray*}
\begin{eqnarray*}
n = 1,2,\ldots
\end{eqnarray*}
\begin{eqnarray*}
\tilde{D}(x) =  \tilde{q}^{-\frac{1}{4}}(x)
\frac{d^2}{dx^2} \tilde{q}^{-\frac{1}{4}}(x) +
\tilde{q}^{-\frac{1}{2}}(x) \delta(x)
\end{eqnarray*}
\begin{eqnarray*}
C(\hbar) = \sum_{n\geq{0}} C_{n} \left(-\frac{\sigma{\hbar}}{2}
\right)^{n}
\end{eqnarray*}
where a choice of a point $x_{0}$ and constants
$C_{k},k=1,2,\ldots$, is arbitrary.  However, for the particular
$\chi_{k}$ (as defined by (\ref{4}), for example) this choice is of course
definite (if $x_{0}$ is given by the lower limit of the
integrations in the expansion (\ref{4}) then $C(\hbar)\equiv 1$).
Nevertheless, even in such cases the choice of $x_0$ can be
arbitrary. Only the constants $C_k$ accompanied to the
choice are definite depending on the choice \cite{7}.  

The representation (\ref{11}) is standard in a sense that any
other one can be brought to (\ref{11}) by redefinitions of the
constants $C_{k}$.  Therefore, any semiclassical expansion can
be uniquely given by fixing $x_{0}$ and the constants $C_{k}$.

And conversly, multiplying a given semiclassical expansion by an
asymptotic series as defined by the last series in (\ref{11})
with other constants $C_{k}, k=1,2,\ldots$, one can obtain any
other semiclassical expansion.

We have mentioned above that the semiclassical series for $\chi_k$
is Borel summable for $x$ staying in the sector $S_{k}$ where $\chi_k$
is defined. In fact it is as such at least inside a circle
$Re({\hbar^{-1}})^* =(2R)^{-1}$ of the $\hbar$-plane satisfying
sufficient conditions of the Watson-Sokal-Nevanlinna (WSN) theorem \cite{10}.

Construct now a new semiclassical series by multiplying the one
for $\chi_k$ by a $\hbar$-dependent constant $C(\hbar)$ with an
{\it analytic} behaviour at $\hbar = 0$.  Expand $C(\hbar)$ into
a power series in $\hbar$, the latter being simultanuously an
{\it asymptotic} expansions for the constant.  Multiply with this
power series the corresponding semiclassical expansion for
$\chi_k$.

A resulting semiclassical series can be now Borel resummed
leading us again to another solution to SE. However, this new
solution can have now two representions: the one being the
solutions (\ref{3}) multiplied by $C(\hbar)$, and the second
being a solution provided by the performed Borel resummation
i.e. there is no a priori a necessity for these two
representations to coincide.

This is exactly what is observed when a change of variable in SE
is performed.

\section{Change of variable as Borel resummations }

Consider therefore a change of variable in (\ref{7}) putting
$y=y(x)$ and assuming $y^{\prime}(x)$ to be meromorphic.
Such a change of variable preserves the SE (\ref{7}) if
simultanuously we make a substitution: $\Phi(y(x)) \equiv
y^{\prime \frac{1}{2}}(x)\Psi(x)$ and $Q(y)$ corresponding to
$\Phi(y)$ in its Schr\"odinger-like equation is given by:

\begin{eqnarray}
{y^{\prime}}^{2}(x)Q(y(x)) = q(x) - \hbar^{2}
\left( \frac{3}{4}\frac{{y^{\prime{\prime}}}^{2}(x)}{{y^{\prime}}^{2}(x)} - 
\frac{1}{2}\frac{y^{\prime{\prime}\prime}(x)}{y^{\prime}(x)} \right)
\label{12}
\end{eqnarray}

Therefore, the above change of variable provide us with a new potential 
differing from the old one by the term which depends totaly on $y(x)$. 
It follows from the form of this term that since $y^{\prime}(x)$ is assumed 
to be meromorphic this dependence can introduce to the new potential at most
second order poles not cancelling the ones of the original potential
$V(x)$ if the latter poles do not depend on $\hbar$. It then follows
further that the new second order poles can introduce to the corresponding
SG additional sectors and SL's not cancelling the old ones built around
the old infinite points of the actions. The old sectors of course change 
their boundaries and enviroments (having possibly as their neighbours 
some new sectors).

Consider now therefore the old sector $S_k$ and its new modified form
$\tilde{S}_k$. Both the sectors have a common part containing $x_0$ 
at its boundary. Using $\Phi(y)$ and $Q(y)$ we can construct
in the $\tilde{S}_k$ a solution $\tilde{\Psi}_{k}(x)$ to SE (\ref{7}). 
Namely, we have:

\begin{eqnarray}
\tilde{\Psi}_{k} = 
\left({y^{\prime}}^{2}\tilde{Q}(y(x)) \right)^{-\frac{1}{4}}
e^{\textstyle {\frac{\sigma}{\hbar}
\int_{x_{i}}^{x}\sqrt{y^{\prime{2}}(\xi)\tilde{Q}(y(\xi))}d\xi}}
\tilde{\chi}_{k}(y(x))
\label{13}
\end{eqnarray}
\begin{eqnarray*}
k=1,2,\ldots
\end{eqnarray*}
where $\tilde{\chi}_{k}(y)$ is constructed according to
(\ref{4}) - (\ref{6}) by making there substitutions: \\
$x{\rightarrow}y(=y(x))$, $\delta(x)\rightarrow\tilde{\delta}(y)$,
$\tilde{q}(x)\rightarrow\tilde{Q}(y)$, $\omega(x)\rightarrow
\tilde{\omega}(y)$,
$W(x)\rightarrow\tilde{W}(y)$ and $x_{0}{\rightarrow}y_{0}(=y(x_0))$.

Note that the new second order poles introduced to (\ref{13})
by $y^{\prime}(x)$ being not present in the original potential
$V(x)$ are not real singularities of $\tilde{\Psi}_{k}(x)$.
They are only singularities of the representation (\ref{13}).

To the solution (\ref{13}) there correspond a domain $\tilde{D}_{k}$
(an obvious analogue of $D$ given by the inequality (\ref{8}))
in which the solution has the same properties $1^{0}, 2^{0}$ above as the
previous ones defined by (\ref{3})-(\ref{6}). In particular
the solutions (\ref{13}) is Borel summable to itself in
$\tilde{S}_{k}$

Let us note further that because the sectors $S_{k}$ and $\tilde{S}_{k}$
have a common part with $x_{0}$ at its boundary then the solutions
(\ref{3}) and (\ref{13}) defined in the corresponding sectors
have to coincide with each other up to a muliplicative constant
$C_{k}$ i.e.
\begin{eqnarray}
\tilde{\Psi}_{k}(x) = C_{k}(\hbar)\Psi_{k}(x)   &   k = 1,2,\ldots
\label{14}
\end{eqnarray}
with $C_{k}(\hbar)$ given by
\begin{eqnarray}
C_{k}(\hbar) = exp \left[ \sigma{\hbar}\int_{x_{i}}^{x_{0}}
\frac{\tilde{\delta}(x)-f(x)}{\sqrt{\tilde{q}(x)} + \sqrt{q(x) +
\hbar^2 \tilde{\delta}(x)
- \hbar^{2}f(x)}}dx \right]
\label{15}
\end{eqnarray}
where
\begin{eqnarray}
f(x) =
\frac{3}{4}\frac{{y^{\prime{\prime}}}^{2}(x)}{{y^{\prime}}^{2}(x)} -
\frac{1}{2}\frac{y^{\prime{\prime}\prime}(x)}{y^{\prime}(x)}
\label{16}
\end{eqnarray}
The coefficient $C_{k}$ was calculated by taking a limit $x
\rightarrow x_{0}$ on both sides of (\ref{14}).

From (\ref{14}) and (\ref{15}) we get the following relation
between $\tilde{\chi}_{k}$ and $\chi_{k}$:
\begin{eqnarray}
\tilde{\chi}_k(x)=
\left(1+\hbar^{2}\frac{\tilde{\delta}(x)-f(x)}{\tilde{q}(x)}
\right)^{\frac{1}{4}}
\exp\left[-\sigma{\hbar}\int_{x_0}^{x}
\frac{\tilde{\delta}(\xi)-f(\xi)}{\sqrt{\tilde{q}(\xi)}+
\sqrt{\tilde{q}(\xi)+\hbar^2\tilde{\delta}(\xi)
-\hbar^{2}f(\xi)}}d\xi \right]\chi_{k}(x)
\label{17}
\end{eqnarray}

Note that the two factors in (\ref{17}) staying in front of $\tilde{\chi}_{k}$
are holomorphic with respect to $\hbar$ at $\hbar=0$.  

We shall now show that the solution (\ref{13}) as well as its
$\tilde{\chi}_{k}$-function are just the Borel sums of the corresponding
semiclassically expanded right hand sights in (\ref{14}) and
(\ref{17}), respectively.

This is an immediate consequence of the holomorphicity of the
coefficient $C_{k}(\hbar)$ and of the two factors in (\ref{17}) at
$\hbar=0$ due to which their semiclassical expansions coincide
with their convergent power series expansion in $\hbar$.
Therefore, due to our earlier discussion the WSN conditions
for Borel summability of the semiclassical series emerging from
RHS in (\ref{14}) and (\ref{17}) are satisfied and
$\tilde{\Psi}_k(x)$ and $\tilde{\chi}_k(x)$ are obtained by
taking these Borel sums.  

\section{Change of variable and exactness of JWKB quantization
formulae}

 The last result can be done even more appealing by
using the following exponential representation for $\tilde{\chi}_k(x)$ and
${\chi}_k(x)$:
\be\label{18}
\tilde{\chi}_k(x) = \exp\left(\int_{x_0}^x \tilde{\rho}_k (\xi)d\xi \right) &,
& {\chi}_k(x) = \exp\left(\int_{x_0}^x {\rho}_k (\xi)d\xi \right)
\ee
so that
\be\label{19}
\tilde{\rho}_k(x) = \frac{\tilde{\chi}_k^{\prime}}{\tilde{\chi}_k}, &
{\rho}_k(x) = \frac{\chi_k^{\prime}}{\chi_k}
\ee
and the relation (\ref{17}) takes the form:
\begin{eqnarray}
\tilde{\rho}_k(x)=\rho_k(x)-\sigma{\hbar}
\frac{\tilde{\delta}(x)-f(x)}{\sqrt{\tilde{q}(x)}+\sqrt{q(x)+
\hbar^2 \tilde{\delta}(x)-\hbar^{2}f(x)}}+ \nn \\
+\frac{\hbar^2}{4} \frac{\tilde{q}(x)}{q(x)+
\hbar^2\tilde{\delta}(x)-\hbar^{2}f(x)}\left(
\frac{\tilde{\delta}(x)-f(x)}{\tilde{q}(x)}\right)^{\prime}
\label{20}
\end{eqnarray}

It follows from (\ref{19}) that both $\tilde{\rho}_k(x,\hbar)$ and
$\rho_k(x,\hbar)$ are Borel summable and from (\ref{20}) that their
Borel transforms differ by a function holomorphic on the whole
Borel plane if both the functions $f(x)$ and $\tilde{\delta}(x)$ are 
$\hbar$-independent. In the latter case it is clear that one cannot 
find such $f(x)$ (the form of $\tilde{\delta}(x)$
has to follow from this of $f(x)$) to cause $\tilde{\rho}_k(x,\hbar)$ to 
disappear i.e. one cannot be left in $\tilde{\Psi}_k(x)$ with its first
two JWKB facors only.
This is because $\tilde{\rho}_k(x,\hbar)$ is singular at $\hbar=0$. 
However, making $f(x)$ to be also $\hbar$-dependent but choosing it 
holomorphic at $\hbar=0$ we can achieve a result when the first $n$ terms of 
the semiclassical expansion of $\tilde{\rho}_k(x,\hbar)$ vanish. The latter 
is possible globally (i.e. independently of $k$) since the semiclassical 
expansions of $\tilde{\rho}_k(x,\hbar)$ are $k$-independent (i.e. do not 
contain any inegration on the $x$-plane, see for example \cite{6}). 
One of our earlier paper is just a good illustration of this possibility 
\cite{6} (see also a comment below).  However, to achieve the goal of 
vanishing $\tilde{\rho}_k(x,\hbar)$ we have to use $f(x,\hbar)$ being 
singular at $\hbar=0$ and therefore being
expected to satisfy all the necessary conditions of
Watson-Sokal-Nevanlinna theorem to be Borel summable. In such a
case $f(x,\hbar)$ becomes, similarly to $\tilde{\rho}_k(x,\hbar)$, sector 
dependent i.e. within the class of the Borel summable functions there is 
no possibilty to define a \underline{global} $y(x,\hbar)$ which when used 
as a variable transformation defining $f(x,\hbar)$ provides us 
with $\tilde{\rho}_k(x,\hbar)$ deprived of its $\tilde{\chi}_k$-factor for
all $k$ simultaneously. In a more obvious way one
can conclude this from (\ref{20}) putting there $\tilde{\rho}_k(x,\hbar)$
 equal to zero and then
treating the equation obtained in this way  as the differential
one for $f(x,\hbar)$ where $\rho_k(x,\hbar)$ is given. However, for
any two different $k$'s there are two different $\rho_k(x,\hbar)$'s
and in consequence two different solutions for $f(x,\hbar)$ have to emerge.

Summarizing the above discussion we can conclude that the effect
of variable changing leading us to the solutions (\ref{13}) can be
obtained also as a result of Borel resummations of the standard
semiclassical expansions for the solutions (\ref{4}) multiplied by a
suitaibly chosen $\hbar$-dependent constants.

A choice of the
constants in (\ref{14}) can be even done in such a way to produce
simultaneously fundamental solutions for which the series in (\ref{5})
start with an arbitrary high power of $\hbar$ \cite{6}. Such a choice
corresponds to a total effect of repeating changes of variable
when for each subsequent Schroedinger-like equation a new
independent variable is the action i.e.  ${y^{\prime}}^2 (x)=
\tilde{q}(x,\hbar)$. The 'lacking'
powers of $\hbar$ are collected then in  $({y^{\prime}}^2 (x,\hbar)
\tilde{Q}(x,\hbar))^{\frac{1}{4}}$ and in the corresponding
exponential factors of the solutions (\ref{4}). These two factors are
then the sources of new JWKB approximations generalizing the
conventional ones \cite{6}.  

Nevertheless, as it follows from the
above discussion, there is no such a choice of the constants $C_k$
which could cause all the corresponding $\tilde{\chi}_k$'s to be reduced to
unity if all the constants as given by (\ref{15}) are to be defined
only by one global $f(x,\hbar)$ given on its own by some $y(x,\hbar)$
realizing the underlying change of the $x$-variable.

Some basic conclusion of the latter statement for the possibility to get
the exact JWKB formula for energy level quantization is the
following.  

Consider a quantization of $1$-dim quantum systems
with the help of the fundamental solutions (it has been
described in many of our earlier papers \cite{1,13,14,16}). Let us
limit the problem to the case when after a change of the
$x$-variable there are only two real turning points $x_1$, $x_2$
of ${y^{\prime}}^2 (x,\hbar)\tilde{Q}(x,\hbar)$
whilst the rest of them are complex and conjugated pairwise
(we assume ${y^{\prime}}^2 (x,\hbar)\tilde{Q}(x,\hbar)$ and E to be real).
We assume also that the problem has
been limited to a segment $z_1 \leq x \leq z_2$ at the ends of which
${y^{\prime}}^2 (x,\hbar)\tilde{Q}(x,\hbar)$ has poles.
In particular we can push any of $z_{1,2}$ (or both of them) to 
$\mp \infty$ respectively.  

To write the corresponding quantization condition
for energy $E$ and to handle simultaneously the cases of second
and higher order poles we assume $z_1$ to be the second order pole
and $z_2$ to be the higher ones.  

It is also necessary to fix to
some extent the closest enviroment of the real axis of the $x$-
plane to draw a piece of SG sufficient to write the quantization
condition. To this end we assume $x_3$ and $\bar{x}_3$ as well as 
$x_4$ and $\bar{x}_4$
to be another four turning points and $z_3$ and $\bar{z}_3$ another two
second order poles of ${y^{\prime}}^2 (x,\hbar)\tilde{Q}(x,\hbar)$ closest 
the real axis. Then a possible piece of SG can look as in Fig.1.  

There is no a unique way of
writing the quantization condition corresponding to the figure.
Some possible three forms of this condition can be written as
\cite{8}: 
\be\label{21}
\exp\ll[\frac{\sigma}{\hbar}\oint\limits_K
({y^{\prime}}^2 \tilde{Q}(x,\hbar))^\fr dx \right]=
-\frac{\chi_{1\to 3}(\hbar)\chi_{2\to
\bar{3}}(\hbar)}{\chi_{1\to
\bar{3}}(\hbar)\chi_{2\to 3}(\hbar)}
=-\frac{\chi_{1\to 4}(\hbar)\chi_{2\to
\bar{3}}(\hbar)}{\chi_{1\to\bar{3}}(\hbar)\chi_{2\to 4}(\hbar)}
\ee
and $\chi_{k\to j}(\hbar) k,j=1,2,3,4$\ are calculated for $x\to z_j$. 
The closed integration path $K$ is shown in Fig.1.
In the figure the paths $\gamma_{1\to 3},\; \gamma_{2\to 3}$,
etc., are the integration paths in the formula (\ref{4})
whilst the wavy lines designate corresponding cuts of the
$x$--Riemann surface on which all the FS are defined.

\vskip 12pt
\begin{tabular}{c}
\psfig{figure=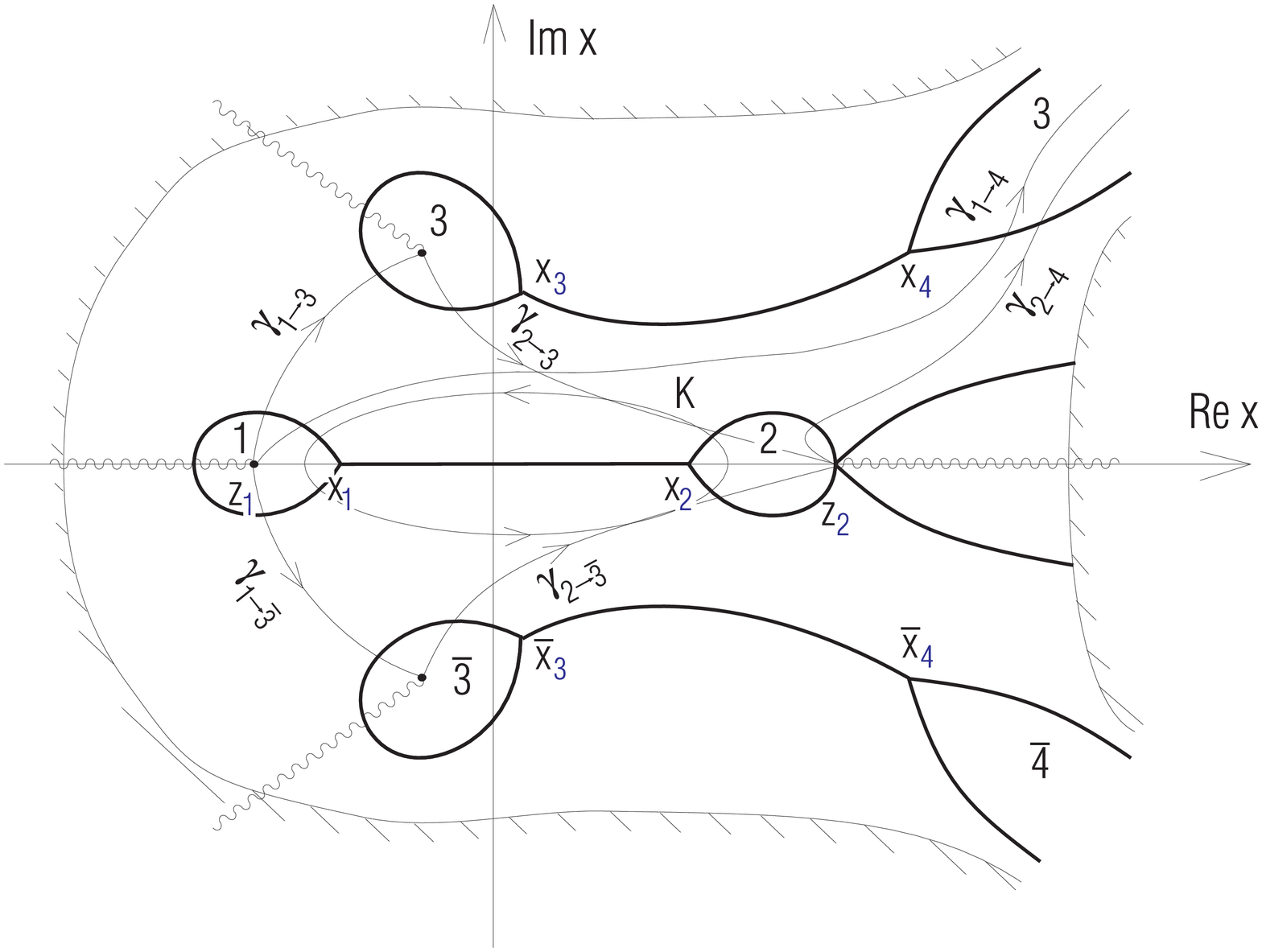, width=12cm} \\
Fig.1  The SG corresponding to general quantization rule \mref{21}
\end{tabular}

\vskip 12pt

The condition (\ref{21}) is $exact$. Its LHS has just the JWKB form. If we
substitude each $\chi_{k\to j}(\hbar)$ in (\ref{21}) by unity 
(which these coefficients
approach when $\hbar \to 0$) we obtain the well- known JWKB quantization
rule which in general is only an approximation to (\ref{21}).

Now, since there is no an $x$-variable transformation $y(x,\hbar)$ by which
all $\chi_{k\to j}(\hbar)$ in (\ref{21}) could become simultaneously equal 
to unity the RHS of (\ref{21}) cannot be reduced to unity by any such 
$y(x,\hbar)$ i.e. the JWKB
formula provided in this way by (\ref{21}) is always only an
approximation. Some additional symmetry conditions have to be
satisfied by the initial $q(x)$ to provide us with such an exact
JWKB formula \cite{11}.

\section{Conclusions}

To conclude we have shown that the Borel summable  fundamental solutions
to SE can be modified by appropriate Borel resummations of the
latter multiplied by properly chosen $\hbar$-dependent constans.
Sometimes effects of such resummations can be recognized as a
proper change of variable in SE. But the latter can always be
considered as an effect of such resummations. This justifies
certainly all the improvements and sometimes exact results
provided by the change-of-variable procedure applied in JWKB
calculations. The latter possibility (i.e. the exact results),
however, can realize only due to particular properties of
considered potentials reflected in global structures of their
respective Stokes graphs \cite{11}.

\section*{Appendix}

The Maslov method is formulated for an arbitrary linear
partial differential equation (LPDE) having as its semiclassical
partner a dynamical system with a finite number of degrees of
freedom \cite{13}. Maslov's semiclassical theory of solutions to the
corresponding LPDE is developed on the 'classical' objects known
in the classical mechanics as Lagrangian manifolds \cite{13,14}.
Limited to the one-degree-of-freedom case and to the $1$-DSE the
Lagrangian manifolds are nothing but the $1$-dim classical
trajectories in the corresponding $2$-dim phase space. Exact
solutions to the stationary Schroedinger equation having
particular Dirac forms (\ref{4}) can be naturally redefined to live on
the Lagrangian manifold (LM) corresponding to a given energy.
However, to cover by such a description the whole coordinate
domain which the corresponding wave fuctions are defined on, the
imaginary time evolution of the classical equations of motion
has also to be switched on to take into account so called
'classically forbidden regions'. The emerging LM contains then
branches corresponding to the real time motions (performed in
classically allowed regions) as well as to the imaginary ones
with the imaginary part of the momentum in the latter case
playing the role of the classical momentum. Of course, the
semiclassical conditions for the considered global wave function
are the following: it has to vanish exponentially when $\hbar \to 0$ in the
classically forbidden regions and to oscillate in the
classically allowed ones.

Unfortunately, the Dirac
representation of these solutions considered as functions of the
coordinate cannot be defined globally on the above LM being
singular at points where the manifold branches making impossible
a matching procedure of the solutions defined on different
branches. These singular points are called in general the
caustic ones but in the $1$-dim case they are known as turning
points. Maslov and Fedoriuk's remedy to solve this arising
'connection problem' is to change the coordinate variable around
such points into the corresponding momentum i.e. to change the
coordinate representation of the wave function into the momentum
one preserving the Dirac form of the solution. Assuming the wave
function to be normalized, its latter representation can be
given formally by the Fourier transformation of the former. In
the new representation the wave function is then regular at the
coordinate turning points of LM (being on the other hand
singular at the emerging momentum turning points). The invers
Fourier transformation considered close to a coordinate turning
point provides us again with the solution in the coordinate
representation given on both the sides of the chosen coordinate
turning point.  As we have mentioned above the semiclassical
limit condition for the latter solution is of course to vanish
exponentially (when $\hbar \to 0$) on one side of the turning point and to
oscillate on the other.  This condition determines the way the
local solutions determined on both the sides of each turning
point and having the Dirac form are to be matched.

The above idea of matching the solutions on different branches the
Lagrangian manifold does not seem to be effective for the exact
solutions to SE but it becomes as such when the solutions in
their Dirac forms are substituted by their corresponding
semiclassical series. This is in fact the subject of the
original approach of Maslov and collaborators. Namely, in such a
case the classically forbidden parts of the solutions
disappeared completely (being exponentially small) and the
remaining ones are then given uniquely on the classically
allowed branches of LM. The matching procedure connects then
only two oscillating solutions separated by the corresponding
turning point. The underlying Fourier transformation becomes
then effectively a point transformation determining the
connection. As it is well known \cite{13} such an semiclassical wave
function continued through a turning point on LM changes its
phase by $\pm 1$. (These changes are controlled in general by so
called Maslov indeces). Synthetically the whole operation is
performed with the help of the Maslov canonical operator \cite{13}.

It is easy to note however that the necessity to use Fourier
transformation diappears if there are possibilities to avoid
somehow turning (caustic) points on the way the wave function is
continued on. This can be achieved for example by enlarging the
number of dimensions the problem is formulated in. The
complexification of the problem is the one of such ways to be
used \cite{15}. In the $1$-dim case this can be done effectively and
without appealing directly to the semiclassical series
expansions by defining the problem on the complex coordinate
plane and utilizing the notions of Stokes graphs and fundamental
solutions. In comparison with Maslov's approach the complex
coordinate plane (in fact the latter is rather a Riemann
surface) corresponds to the complex Lagrangian manifold endowed
with the coordinate charts collected of all canonical domains
defined by the corresponding Stokes graph. To each canonical
domain a fundamental solution is attached having the
corresponding domain as the maximal one where its semiclassical
expansion as given by (\ref{11}) is valid. There is no necessity to
construct and to use the Maslov canonical operator to continue
(analytically) the fundamental solutions and to match them in
any domain of the plane. The Maslov indeces gained by the
fundamental solutions on the way of their analytical
continuations are provided by crossed cuts of the corresponding
Riemann surface. Therefore using the fundamental solution method
in the $1$-dim problems is completely equivalent to the
corresponding Maslov one in the semiclassical regime of the
problem but it has many obvious advantages over the latter with
their use as the exact solutions to SE being the first one.
Other important properties of the method have been mentioned and
used in the main body of this as well as other papers \cite{6,7,8,9,11}.

\end{document}